\documentclass{emulateapj}
\usepackage{apjfonts,epstopdf}

\journalinfo{Astrophysical Journal Letters, in press}
\slugcomment{}

\shorttitle{AGN clustering in ECDF-S at $z\sim3$}
\shortauthors{Francke et al.}

\begin{document}

\title{Clustering of Intermediate Luminosity X-ray selected AGN at $z\sim3$\footnotemark[6]}

\author{ Harold Francke\footnotemark[1], Eric
  Gawiser\footnotemark[4], Paulina Lira\footnotemark[1],
  Ezequiel Treister\footnotemark[5], Shanil Virani\footnotemark[2],
  Carie Cardamone\footnotemark[2], C.M. Urry\footnotemark[3],
  Pieter van Dokkum\footnotemark[2] and Ryan Quadri\footnotemark[2]}

\footnotetext[1]{Departamento de Astronom\'{\i}a, Universidad de
Chile, Casilla 36-D, Santiago, Chile.}
\footnotetext[2]{Department of
Astronomy, Yale University, PO Box 208101, New Haven, CT 06520.}
\footnotetext[3]{Yale Center for Astronomy \& Astrophysics and
 Department of Physics, Yale University, P.O. Box 208121, New Haven, CT 06520.}
\footnotetext[4]{Department of Physics and Astronomy, Rutgers, 136
 Frelinghuysen Road, Piscataway, NJ 08854-8019}
\footnotetext[5]{European Southern Observatory, Santiago, Chile}
\footnotetext[6]{This work is based on observations made with the 6.5
m Magellan-Baade telescope, a collaboration between the Observatories
of the Carnegie Institution of Washington, University of Arizona,
Harvard University, University of Michigan, and Massachusetts
Institute of Technology, and at Cerro Tololo Inter-American
Observatory, a division of the National Optical Astronomy
Observatories, which is operated by the Association of Universities
for Research in Astronomy, Inc. under cooperative agreement with the
National Science Foundation.}

\email{hfrancke@das.uchile.cl}

\defcitealias{adelberger05cons}{A05b}
\defcitealias{shenetal2007}{S07}

\begin{abstract} 

  We present the first clustering results of X-ray selected AGN at
  $z\sim3$. Using Chandra X-ray imaging and UVR optical colors from
  MUSYC photometry in the ECDF-S field, we selected a sample of 58
  $z\sim3$ AGN candidates. From the optical data we also selected 1385
  LBG at 2.8 $< z <$ 3.8 with R $<$ 25.5.  We performed
  auto-correlation and cross-correlation analyses, and here we present
  results for the clustering amplitudes and dark matter halo masses of
  each sample.  For the LBG we find a correlation length of $r_{0,LBG}
  = 6.7 \pm 0.5$ Mpc, implying a bias value of 3.5$\pm$0.3 and dark
  matter (DM) halo masses of log($M_{min}$/$M_{\odot}$) =
  $11.8\pm0.1$. The AGN-LBG cross-correlation yields $r_{0,AGN-LBG} =
  8.7\pm1.9$ Mpc, implying for AGN at 2.8 $<z<$ 3.8 a bias value of
  5.5$\pm$2.0 and DM halo masses of log($M_{min}$/$M_{\odot}$) =
  $12.6_{-0.8}^{+0.5}$.  Evolution of dark matter halos in the Lambda
  CDM cosmology implies that today these $z\sim3$ AGN are found in
  high mass galaxies with a typical luminosity $7_{-2}^{+4} L^{*}$.

\end{abstract}


\section{INTRODUCTION}
\label{sec:intro}
There is a wealth of evidence that nuclear supermassive black holes
play a significant role in the process of galaxy formation and
evolution.  This has become evident in the past few years with the
discovery of correlations between the properties of the massive black
holes, the stellar systems that host them, and their dark matter halos
(\citealp{magorrianetal98},\citealp{gebhardtetal00},
\citealp{ferrarese02}).

Luminous quasars have been studied in great detail, with systematic
spectroscopic studies by the Sloan Digital Sky Survey and the 2dF
Galaxy Redshift Survey.  These objects are rare and represent the
bright end of the active galactic nuclei (AGN) luminosity function.
In order to fully understand the link between the growth of
supermassive black holes (SMBH) and galaxy evolution, we need to study
young, high-redshift galaxies hosting AGN with more typical
luminosities.

One of the basic properties of galaxy populations is their clustering
strength, but there are few constraints on this quantity for fainter,
high-redshift AGN. Furthermore, there are important disagreements in
the literature: e.g. galaxy-AGN cross-correlation measurements at
$z\sim3$ by \citet{adelberger05cons} \citepalias{adelberger05cons},
imply a bias factor of $3.9\pm3.0$ for luminous AGN (UV luminosities
between -30$<M_{1350}<$-25), while recent work by \citet{shenetal2007}
\citepalias{shenetal2007} suggests a value of $9.1\pm0.9$ for sources
with similar luminosities at the same redshift.  For galaxies, e.g.,
measurements indicate a strong luminosity dependence in the clustering
length \citep{giavalisco2001}. On the other hand, for AGN there are a
handful of claims that this is not the case (\citealp{croometal2005},
\citealp{myersetal2006}, \citealp{adelbergers05}). Determining the
bias with accuracy puts important constraints on models of AGN
formation and evolution, (e.g., \citealp{lidzetal06},
\citealp{hopkinsetal07}) but requires spanning a broad range in
luminosity and obscuration level. This is challenging using optical
color selection plus spectroscopy, because the low fraction of AGN
(3\%) found among these candidates \citep{steideletal02} demands huge
amounts of spectroscopy time.  Using X-ray detection plus spectroscopy
(e.g., \citealp{szokolyetal2004}) provides a more efficient and
unbiased selection of AGN, since their surface density is higher in
X-rays than in optical images and obscuration effects are much less
important. However, the unrestricted redshift range sampled by this
method makes spectroscopic follow-up highly inefficient, and often
prevents confirmation of dimmer, high-z AGN (R$>$24). Hence it is
difficult to obtain AGN samples suitable for clustering studies.
There have been several measurements of the spatial correlation
function of X-ray selected AGN at $z<1$ (\citealp{mullisetal2004},
\citealp{gillietal04}, \citealp{basilakosetal2004},
\citealp{basilakosetal2005}, \citealp{yangetal2006},
\citealp{miyajietal2007}), with samples ranging from 200-500
sources. At higher redshifts, the statistics are much poorer, and
consist of purely optically selected AGN. In this work, we constrain
the clustering strength of an AGN sample at $z\sim3$ jointly selected
by optical and X-ray photometry.  We determine whether these
sources cluster more or less than non-active galaxies at this
redshift, and discuss their present-day descendants.

We assume a $\Lambda$CDM cosmology consistent with WMAP results
\citep{spergeletal2007} with $\Omega_m=0.3, \Omega_\Lambda=0.7$, $H_0
= 70$ km s$^{-1}$ Mpc$^{-1}$ and $\sigma_{8}=0.8$. All quantities are
comoving: correlation lengths scale as $h_{70}^{-1}$, number densities
as $h_{70}^{3}$ and halo masses as $h_{70}^{-1}$.

\section{OBSERVATIONS}
\label{sec:obs}

The MUSYC survey was optimized to study galaxies at $z\sim3$, with
imaging depths down to the spectroscopic limit, $U$, $B$, $V$, $R$
$\sim$ 26 \citep{gawiseretal06b}. In the ECDF-S field, in
particular, the 5$\sigma$ limiting (AB) magnitudes achieved in these
bandpasses are 26.0, 26.9, 26.4 and 26.4 respectively. The main MUSYC
catalog is based on the sources detected on the combined BVR image,
and aperture photometry is performed on each filter in those
positions (see \citealp{gawiseretal06a}).

The deep Chandra observations of this field have produced four X-ray
catalogs: \cite{giacconietal02}, \cite{alexanderetal03},
\cite{lehmeretal05b}, and \cite{viranietal06}. The first two comprise 1
Ms of exposures inside the central region (CDF-S proper), covering an
area of $\approx$0.1 deg$^2$ (PI R. Giacconi). The
\citeauthor{alexanderetal03} catalog has reported flux limits of $5.2
\times10^{-17}$ erg cm$^{-2}$ s$^{-1}$ and $2.8\times10^{-16}$ erg
cm$^{-2}$ s$^{-1}$ in the soft (0.5-2 keV) and hard (2-8 keV) bands,
respectively. The last two catalogs come from the four $\approx$250 ks
pointings that cover an area of $\approx$0.3 deg$^2$ around the former
field (PI N.Brandt). Limiting X-ray fluxes in the extended region are
$1.1\times10^{-16}$ erg cm$^{-2}$ s$^{-1}$ in the soft and
$6.7\times10^{-16}$ erg cm$^{-2}$ s$^{-1}$ in the hard bands,
correspondingly.

The MUSYC spectroscopic follow-up program carried out with
Magellan/Baade+IMACS has yielded 280 successful identifications of z
$>$ 2 galaxies. Data were obtained with a resolution of R=640 (470
km/s at 5000\AA) and slitlets of 1.2$''$ (P. Lira et al., in
prep). The broad wavelength coverage provided by IMACS, between 4000
and 9000\AA, allows the detection of Ly$\alpha$ and C~{\small IV}
1549, the two most prominent ultraviolet AGN emission lines, at
redshifts 2.3 $< z <$ 4.0.

\section{AGN AND LBG SAMPLES}
\label{sec:sample}

The full set of unique X-ray counterparts is taken from the MUSYC
ECDF-S X-ray catalog (Cardamone et al. 2007, submitted to ApJ), constructed
joining the catalogs from \citet{giacconietal02},
\citet{alexanderetal03}, \citet{viranietal06} and
\citet{lehmeretal05b}, using a likelihood procedure
\citep{brusaetal2005} to match sources between the X-ray and optical
(BVR-selected) catalogs.

Fig.~1 shows the $U - V$ versus $V - R$ color-color plot for sources
in the ECDF-S from MUSYC photometry. Lyman break color selection
(`UVR') corresponds to the region outlined in the upper left side
(\citealp{steideletal96},\citealp{gawiseretal06a}).  For this color
selection, $U$-band fluxes are required to be detected at 1-$\sigma$,
and are otherwise set to an upper limit equal to their 1-$\sigma$
error. This avoids interlopers (typically dwarf stars) with uncertain
photometry at the cost of incompleteness in the sample.  Additionally,
$R$ $<$ 25.5 is required to allow for spectroscopic
confirmation. Sources presenting a drop in the $U$ filter due to
intergalactic absorption, along with a blue continuum between the $V$
and $R$ filters that rules out heavily reddened lower-z objects, are
expected to lie at 2.7 $< z <$ 3.7.  For XUVR selection of AGN we
require an X-ray detection with counterpart UVR colors in the same
region and we drop the R magnitude limit and the 1-$\sigma$ $U$-band
requirement, since the extra requirement of X-ray emission already
rules out most dwarf stars. This procedure yields 1385 LBG and 58 AGN
candidates. Unobscured AGN and LBG have somewhat similar UVR colors,
as can be seen in the color-color track in Fig.~1, since these colors
are mainly determined by the intergalactic hydrogen
absorption. Furthermore, obscured AGN are the dominant population
among AGN, and since the SED of the first is dominated by their host
galaxies (\citealp{treisteretal04b}, \citealp{treisterurry05}), we
expect the redshift distribution of our AGN sample to differ mildly
from the LBG distribution. Notice that in the case of significant
obscuration, we don't expect to select AGN hosted by very red
galaxies, since they will probably not be bright enough in the
restframe UV.

\begin{figure}
\epsscale{1.1}
\plotone{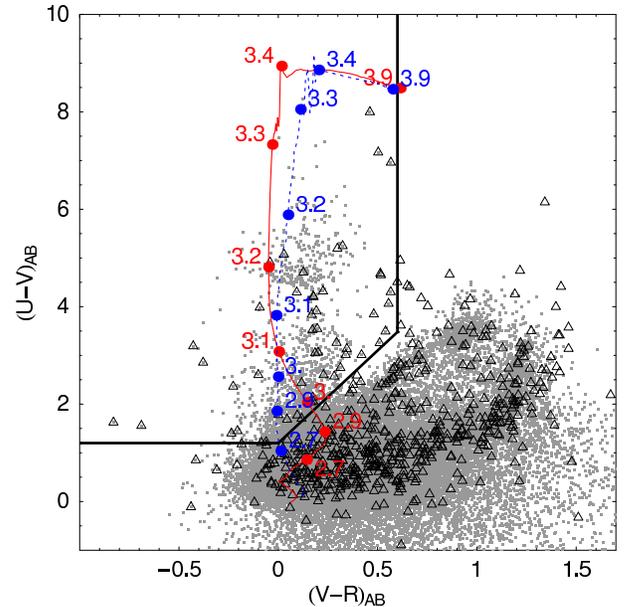}
\caption{$U - V$ versus $V - R$ color-color plot for all the sources
brighter than $R=25.5$ in the MUSYC ECDF-S optical catalog (gray
points) and X-ray counterparts (open triangles). LBG and $z\sim3$ AGN
selection are inside the region marked in solid lines. The color-color
tracks of a template AGN and a LBG as they are redshifted from z=2.0
to 3.9 are shown in the red solid and blue dotted curves,
respectively.
\label{fig:xuvrcolcol}}
\end{figure}

From the spectra obtained over the entire 4-field survey so far, 131
LBG and 30 AGN were identified at the target redshift $\sim$3, and in
the ECDF-S in particular, 31 LBG and 11 AGN have been confirmed.
While AGN in the ECDF-S were directly targeted using a joint X-ray and
optical selection, in the rest of the MUSYC fields AGN have been
discovered serendipitously among the UVR candidates (see Fig. 1).
Fig. 2 shows the redshift histograms of the confirmed objects, which
implies X-ray luminosities between $10^{43}$ and $10^{45}$ ergs/s for
the entire set of AGN candidates. The Lyman-Break (UVR) selection
shows a fraction of low-redshift interloper less than 10\%. In the
spectra of the XUVR-targeted AGN in ECDF-S, we did not observe any
interlopers.

\begin{figure}
\epsscale{1.2}
\plotone{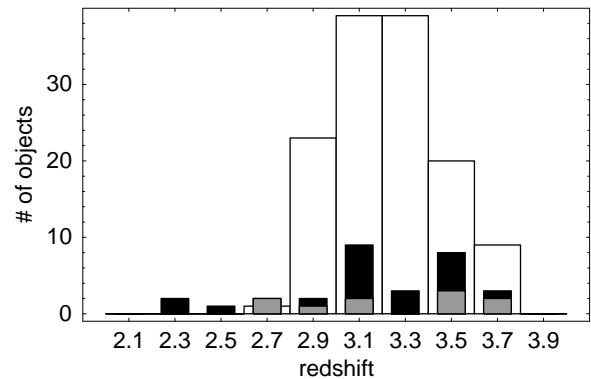}
\caption{ Redshift distribution of all spectroscopically confirmed
LBGs (white bars) and $z\sim3$ AGN (black bars) in MUSYC. The subset
of AGN in ECDF-S is in gray bars.  These distributions have mean and
variances: \={z$_{AGN}$} = 3.18, $\sigma_{z,AGN}$ = 0.39,
\={z$_{LBG}$} = 3.23, $\sigma_{z,LBG}$ = 0.22.
\label{fig:zhist}
}
\end{figure}

\section{CLUSTERING ANALYSIS}
\label{sec:clustering}

We calculate the clustering of AGN via a two-step method, calculating
the autocorrelation of LBG first, and then their cross-correlation
with AGN. The main advantage of this approach is that it improves the
statistics significantly, since the LBG population is much more
numerous and its redshift distribution is similar to the AGN
population sampled here, the AGN-LBG cross-correlation function less
noisy than the AGN autocorrelation function
\citep{kauffmannhaenelt02}.  Notice that the AGN and LBG that
contribute to the cross-correlation function will be those that
spatially overlap. Therefore, our results will reflect the clustering
of AGN in the redshift range 2.7 $<z<$ 3.8.

For the UVR selected sources, we calculated the angular
autocorrelation function using the \citet{landys93} estimator:
\begin{equation}
\widehat{\omega}_{LS}(\theta) = \frac{DD(\theta)-2DR(\theta)+RR(\theta)}{RR(\theta)}
\end{equation}
where DD, DR and RR are the data-data, data-random and random-random
pairs, all normalized to integrate to 1. To estimate the angular
cross-correlation between UVR-XUVR sources, we used the following
cross-correlation estimator \citep{croftetal1999}:
\begin{equation}
\widehat{\omega}_{AGN-LBG}(\theta) = \frac{D_{AGN}D_{LBG}(\theta)}{D_{AGN}R_{LBG}(\theta)} - 1
\end{equation}
where now the data-data pairs are between AGN and LBG, and data-random
between AGN and the optical random catalog. Notice that this estimator
does not use a random catalog for the X-ray sources. The degradation
of the PSF and sensitivity as a function of the off-axis angle in the
Chandra data makes the X-ray survey geometry fairly complicated, but
these resulting systematic uncertainties can be avoided in this
cross-correlation scheme.

The estimates shown above were fitted to $\widehat{\omega}(\theta) =
A\theta^{-\beta} - \mathrm{IC}(A,\beta)$, where the integral
constraint factor IC is included as part of the model and we assumed
the usual power law for the true angular correlation function. We kept
$\beta=0.8$ fixed, since our S/N does not allow us to put significant
constraints on the slope. Fit limits have been set to avoid the
one-halo term regime on small angular scales (30$''$ for LBG and
60$''$ for AGN based on the virial radii of their initially inferred
DM halo masses of 6$\times 10^{11}$ and $10^{13} M_{\odot}$,
respectively) and ending at half the field size on the larger scales,
in order to avoid border effects of the estimator and where the
sampling becomes poor anyway. The fit to the LBG autocorrelation
function gave $A_{LBG}=2.2\pm0.3~\mathrm{arcsec}^{\beta}$ and the fit
to the cross-correlation function
$A_{AGN-LBG}=2.9\pm1.1~\mathrm{arcsec}^{\beta}$, with reduced
$\chi^{2}$ values of 1.24 and 1.21, respectively. The errors in these
parameters were calculated using $\Delta \chi^{2}$ and correspond to
1-$\sigma$ confidence level with 1 parameter.  Fig.~3 shows the
measured angular correlation functions with their corresponding best
fit models. Binning $\omega(\theta)/\sigma_{\omega}(\theta)$ over the
entire fitting range gives estimates for the total signal-to-noise
ratios of these measurements of 10.2 and 3.9, correspondingly. From
montecarlo realizations, we estimated that the probabilities of
obtaining clustering this high from unclustered populations of the
same sizes are $<$0.1\% for the LBG autocorrelation and ~1\% for the
AGN-LBG cross-correlation.

\begin{figure}
\epsscale{1.1}
\plotone{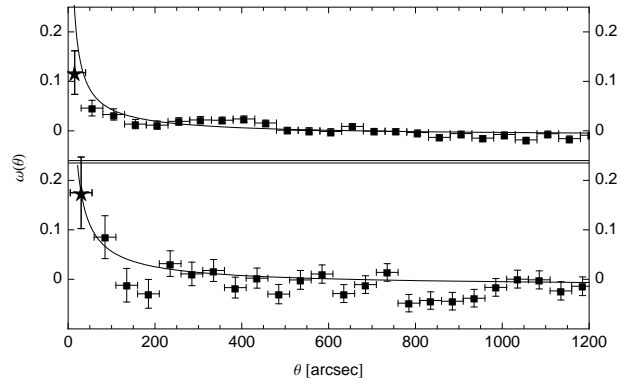}
\caption{Top: Angular autocorrelation function for UVR selected
sources, i.e. Lyman-break galaxies at $z\sim3$. The solid line is the
best fit to the data,
$\omega_{lbg}(\theta)=2.2\pm0.3~\theta^{-0.8}$. The point marked with
a star is not included in the fit and is shown only for
reference. Bottom: Analogous plot for the angular cross-correlation
function with XUVR sources, i.e., AGN at $z\sim3$. The best fit
corresponds to $\omega_{agn-lbg}(\theta)=2.9\pm1.1~\theta^{-0.8}$.
\label{fig:wtheta}
}
\end{figure}

Using smooth fits to the redshift distributions of the confirmed
sources shown in Fig. 2, we deprojected the values obtained for the
angular LBG autocorrelation and LBG-AGN cross-correlation amplitude
using Limber's equation. We obtained spatial correlation lengths of
$r_{0,LBG}=6.7\pm0.5$ Mpc and $r_{0,AGN-LBG}=8.7\pm1.9$ Mpc
respectively, where the correlation length is defined by the usual
fitting form $\xi(r)=(r/r_{0})^{-\gamma}$ for the spatial correlation
function. The random errors introduced in the deprojection are
included in the error budget by approximating the redshift
distributions as Gaussians and propagating the standard errors in
their means and variances.

To calculate the bias factor of the AGN sample in the approximation of
linear, constant bias, we used the elementary relations $b^{2}_{lbg} =
\sigma^{2}_{8,lbg}/\sigma^{2}_{8,dm}$ and $b_{agn}b_{lbg} =
\sigma^{2}_{8,agn-lbg}/\sigma^{2}_{8,dm}$, where $\sigma^{2}_{8,X}$ is
the variance in spheres of radius $8 h_{100}^{-1}$ Mpc in the random
field described by the corresponding auto- or cross-correlation
function. Although $A/\beta$ and $r_0/\gamma$ are highly degenerate
parameter pairs, the bias factors are robust to variations in the
slope $\beta$ (or $\gamma$). Therefore, they are our preferred
quantity for comparison with the literature. The bias, number
densities and masses shown in Table 1 are calculated using the
ellipsoidal collapse model extension of the Press-Schechter formalism
by \citet{shethtormen99} and the extension of the \citet{mow96}
formalism by \citet{shethetal2001}. These quantities were evaluated at
the mean redshifts of the confirmed samples, both very close to
\={z}=3.2.

\begin{deluxetable}{cccc}
\tablecolumns{4}
\tablecaption{Summary of DM halo properties}
\tablehead{
  \colhead{} &
  \colhead{bias} &
  \colhead{log($M_{min}$)} &
  \colhead{$n_{halo}$ (Mpc$^{-3}$)}
}
\startdata
LBG & $3.5\pm0.3$& $11.8\pm0.15$& $9\pm4\times10^{-4}$\\
AGN & $5.5\pm2$& $12.6_{-0.8}^{+0.5}$& $10^{-3}$ - $10^{-6}$
\enddata
\tablecomments{Confidence intervals/ranges are 1-$\sigma$.
Masses are given in logarithm of solar masses.}
\end{deluxetable}

\section{DISCUSSION}
\label{sec:sumdisc}

We have measured the autocorrelation strength of z=3 LBG and the
cross-correlation between these galaxies and AGN selected using both
X-ray and optical data. For the LBG sample, we obtained a bias factor
of 3.5$\pm$0.3. This is somewhat higher than the value 2.8$\pm$0.3
found by \citet{adelbergeretal05a} at $z=3$. \citet{hildebrandtetal07}
obtained a bias value of 3.2$\pm$0.2 for an equivalent LBG population,
consistent with ours.

From the AGN-LBG cross-correlation and the bias calculated for the
LBG, we have deduced a bias factor of 5.5$\pm$2 for our AGN
sample. The active galaxies targeted in this study appear to cluster
more than star-forming galaxies with similar restframe-UV colors. This
is consistent with cosmic downsizing of AGN, implying that typical SMBH
tend to sit in more massive galaxies than the ``normal'' galaxy
population. We need greater statistics to confirm this
result, since the values are consistent within the uncertainties.

\citetalias{adelberger05cons} performed this same calculation in an
optically selected and spectroscopically confirmed sample of 79 AGN
between 1.6 $< z <$ 3.7, dividing the AGN by UV luminosity into bright
(25 sources with -30$<M_{1350}<$-25) and dim (54 sources with
-25$<M_{1350}<$-19) samples. Although \citetalias{adelberger05cons}
does not report the autocorrelation length of the galaxy sample used
to compute these values, we approximate it by averaging the result for
LBG and BX galaxies presented in \citet{adelbergeretal05a}, obtaining
a bias factor of 2.6$\pm$0.3. From this we infer bias factors of
3.9$\pm$3.0 and 4.7$\pm$1.7 for the \citetalias{adelberger05cons}
bright and dim AGN samples. Our AGN sample has an UV magnitude range
between -26 and -20, almost identical to the
\citetalias{adelberger05cons} faint AGN set, and showing the same
clustering strength. In Fig.~4 these results are compared to the
present estimate and to that obtained by \citetalias{shenetal2007} for
$\sim$2250 SDSS quasars at 2.9$<z<$3.5.

To estimate the dark matter halo mass of the typical descendant of the
halos that host these $z \sim 3$ galaxy sets at the present time, we
calculated the mode and width of the conditional probability
distribution of the expected mass $z=0$ (see \citealp{hamanaetal06}
and references therein). We took the median mass of the halo
populations at $z=3.2$ as the representative value, and we report the
bias values corresponding to the maximum likelihood z=0 progenitor,
with $1-\sigma$ uncertainties corresponding to halo masses whose
likelihood is reduced by a factor $exp(-1/2)=0.607$. For the LBG bias
of 3.5 at $z = 3.2$, we obtain a bias of $1.3_{-0.1}^{+0.3}$ at
$z=0$. In the nearby universe \citep{zehavietal2005}, this corresponds
to the clustering of a somewhat massive galaxy, with
$L=2.7_{-0.6}^{+1.9} L^{*}$. On the other hand, our AGN sample, with a
bias of 5.5 at $z=3.2$ would have a typical halo with a bias factor of
$2.0_{-0.3}^{+0.6}$ at present. Extrapolating from
\citet{zehavietal2005}, this corresponds to the clustering of galaxies
having $7_{-2}^{+4} L^{*}$, which at the present time are most likely
located in groups and galaxy clusters.

\begin{figure}
\plotone{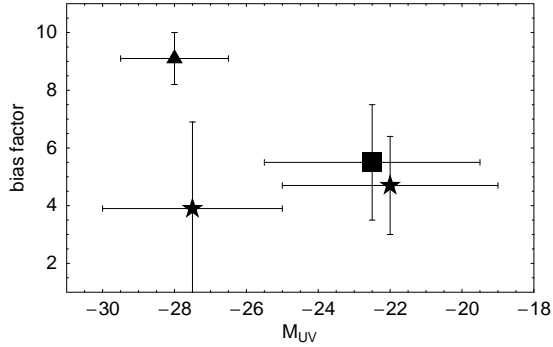}
\caption{ Comparison of AGN bias factors at $z\sim3$ between this work
(square), \citetalias{shenetal2007} (triangle) and
\citetalias{adelberger05cons} (stars).  
\label{fig:biascomp}
}
\end{figure}

This confirms the result found by \citetalias{adelberger05cons} for
faint, optically selected AGN. Improving our understanding of the
AGN-galaxy connection at $z\sim3$ and constraining models such as
\citet{lidzetal06} and \citet{hopkinsetal07} for AGN clustering,
requires resolving the discrepancy in the bias estimates for bright
AGN seen in Fig.~4 improving the statistics. We have show that
targeting high-redshift AGN for clustering analyses can be done very
efficiently by means of deep optical and X-ray imaging, and for that
reason, future surveys with deep X-ray coverage will be ideal for
obtaining large samples of active galaxies in restricted redshift
ranges, suitable for clustering studies. Our method can be applied to
these surveys to obtain enough X-ray-selected AGN to reduce the
current uncertainties.

H.~F. was supported by MECESUP project UCH0118, Andes Fundation
fellowship C-13798, ALMA fellowship 31060003 and Fondecyt project 1040719.

\end{document}